\documentclass{article}

\usepackage{PRIMEarxiv}
\usepackage{amsmath}
\usepackage{algorithmic}
\usepackage{algorithm}
\usepackage[utf8]{inputenc} 
\usepackage[T1]{fontenc}    
\usepackage{hyperref}       
\usepackage{url}            
\usepackage{booktabs}       
\usepackage{amsfonts}       
\usepackage{nicefrac}       
\usepackage{microtype}      
\usepackage{lipsum}
\usepackage{fancyhdr}       
\usepackage{graphicx}       
\graphicspath{{media/}}     

\title{Adaptive Multi-Layer Deployment for a Digital Twin Empowered Satellite-Terrestrial Integrated Network
\thanks{\textit{\underline{Citation}}: 
\textbf{Yihong TAO, Bo LEI, Haoyang SHI, et al., 2024. Adaptive multi-layer deployment for a digital twin-empowered satellite-terrestrial integrated network. Frontiers of Information Technology and Electronic Engineering (former title: Journal of Zhejiang University SCIENCE C (Computers and Electronics), 2010-2014), in press. https://doi.org/10.1631/FITEE.2400327.}} 
}

\author{
  Yi-hong Tao, Bo Lei, Hao-yang Shi, Jing-kai Chen, Xing Zhang \\
  Wireless Signal Processing and Network Laboratory Beijing University of Posts and Telecommunications \\
  Research Institute of China Telecom Co., Ltd., Beijing 102209, China\\
  2018211025@bupt.cn,hszhang@bupt.edu.cn
}

\begin{document}
\maketitle

\begin{abstract}
With the development of satellite communication technology, satellite-terrestrial integrated  networks (STIN), which integrate satellite networks and ground networks, can realize seamless global coverage of communication services. Confronting the intricacies of network dynamics, the diversity of resource heterogeneity, and the unpredictability of user mobility, dynamic resource allocation within networks faces formidable challenges. Digital twin (DT), as a new technique, can reflect a physical network to a virtual network to monitor, analyze, and optimize the physical network. Nevertheless, in the process of constructing the DT model, the deployment location and resource allocation of DTs may adversely affect its performance. Therefore, we propose a STIN model, which alleviates the problem of insufficient single-layer deployment flexibility of the traditional edge network by deploying DTs in multi-layer nodes in a STIN. To address the challenge of deploying DTs in the network, we propose multi-layer DT deployment in a STIN to reduce system delay. Then we adopt a multi-agent reinforcement learning (MARL) scheme to explore the optimal strategy of the DT multi-layer deployment problem. The implemented scheme demonstrates a notable reduction in system delay, as evidenced by simulation outcomes.
\end{abstract}

\keywords{Digital twin; Satellite-terrestrial integrated network; Deployment; Multi-agent reinforcement learning}

\section{Itroduction}
With the development of networks, STIN are new network types that can make full use of ground networks and satellite networks to meet user service requirements \cite{1,2}. According to network environment and service requirements, STIN provide global seamless coverage and diversified mobile access services for various types of mobile terminals, breaking through geographical and environmental limitations. In remote areas, where aircraft and oceangoing ships cannot be covered by ground networks, satellites can provide network services. Systems specializing in commercial mobile satellite communication, exemplified by Iridium and dedicated maritime satellites, offer proficient solutions tailored to maritime operations, emergency response, and personal mobile communication needs. The emergence of substantial commercial low-orbit satellite constellations, notably OneWeb, Starlink, and Space Norway, underscores the pivotal role that low-orbit satellites are poised to play as a robust augmentation to the anticipated 6G ground networks of the future \cite{3,4,5}. 

Satellite networks transmit data and information through satellite signals and possess the capability to encompass diverse geographical territories extensively, including remote areas and places such as oceans that cannot be covered. Terrestrial networks provide more stable and high-speed communication services in cities and densely populated areas, and are also used for data transmission and connectivity with satellite networks. Therefore, academia and industry are promoting the combination of the two, which can achieve communication coverage and connectivity on a global scale, meeting the communication requirements of different regions and needs \cite{38,39,40}. For different application scenarios, STIN need to meet the differentiated of large-scale connection of users. However, the existing independent control of multi-dimensional satellite ground resources prevents many physical resources in heterogeneous nodes from being effectively utilized, which affects the system's ability to provide reliable services for users \cite{6}\cite{47}\cite{48}. A pressing need arises for a revolutionary technology that can significantly bolster network flexibility, streamline decision-making processes, and elevate the proficiency of data management practices. The DT emerging intelligent technology can monitor, control, and optimize the physical entity, and is expected to solve the above problems well \cite{7,8}. 

DT technology can create accurate virtual twin copies of physical objects through digital technology, and the created DT can communication with the object in the real world in real time, and dynamically reflect its state and change. On the one hand, DT technology enables the physical system to be monitored and managed in real time, including the monitoring of system status, fault detection, and early warning. On the other hand, DT technology can help predict the behavior and performance of real systems and optimize them accordingly. By simulating and analyzing the DT data, problems in the actual system can be found and solved, and the efficiency and the system performance can be improved \cite{9,10}. Some work has proposed using DTs to solve the problems existing in STIN. \cite{11} proposed to use DTs to assist in solving the problem of possible loops in the satellite handover process. \cite{12} established a DT copy for the satellite network to alleviate the challenges of overly complex satellite network design, simulation, deployment, and maintenance. \cite{13} ensured the security of wireless communication between the satellite and ground through DT-assisted multi-dimensional domain cooperative precoding. \cite{14} proposed a novel approach utilizing dynamically adapted micro-cloud architecture, grounded in DTs and multi-agent system technology, which was introduced to tackle the challenge of optimizing satellite liaison window task scheduling priorities. \cite{15} established a handover prediction and congestion prediction model for inter-satellite link ISL in the DT network to ensure the quality of satellite communication services. \cite{16} established a DT-driven satellite-terrestrial integrated edge computing network to optimize resource allocation.
 
The premise of the above research is that DTs have been properly deployed in the server, however, the deployment of DTs in the network affect their performance, and building DT models for users requires significant data communication capacity, and computing capacity to process data, so DT deployment is a basic problem that needs further study. As stated in \cite{17}, when considering DT deployment solutions, we should consider the unique requirements of each application, which vary significantly.  These requirements encompass latency sensitivity, the desired level of service experience quality, the specific allocation of computing resources, and the system reliability. On the one hand, the dynamic state in the network may affect DT performance and affect the quality of services. On the other hand, the computing resources required to maintain DTs in the server and ensure low-latency data interaction between physical entities and DTs need to be considered, which makes the DT deployment problem different from traditional placement problems \cite{18,19}. There are relevant studies proposing DT deployment in the edge network to form digital twin edge networks (DITEN), \cite{20} formulated the DT placement challenge as an edge association task and proposed a scheme based on federated learning to reduce communication and computing latency. \cite{21} proposed the use of DTs in the entire mobile edge computing (MEC) network to provide an offloading scheme in DITEN to minimize system latency.  However, with the further development and application of DT technology, there are higher requirements for the deployment of DT models, such as providing remote DT services for users at sea; in this challenging situation, considering the cost and user service experience, it is not feasible to use only a traditional ground network to transmit and process data. Introducing DT technology into the STIN has become a promising solution to achieve global coverage, improve network reliability, and ensure user service quality. Therefore, it is necessary to further study the application of DTs in the STIN, and how to ensure the interaction between physical space and DT space in the highly dynamic and complex heterogeneous network environment. To our knowledge, no previous work has investigated DT deployment in DT-assisted STIN.

We first describe DT technology in the STIN and propose a DT satellite-terrestrial network model. By placing DTs in different nodes in the network, we propose a DT multi-layer deployment problem. Finally, we propose a solution to this problem using the multi-agent reinforcement learning (MARL) algorithm. The primary contributions of research are outlined as follows.

1. We propose a DT architecture for a STIN. DTs can capture user real-time status and monitor and optimize users. We propose to deploy DTs in multi-layer nodes in the network, which alleviates the problem of insufficient flexibility in a single-layer deployment of traditional DITEN.

2. To address the challenges of deploying DTs in the network, we innovatively formulate the DT multi-layer deployment problem in a STIN to reduce system delay, ensure the interaction between users and DTs, and improve the user experience of DT services.

3. We propose an algorithm based on MARL that formulaically solves the proposed DT deployment problem by considering the DT deployment strategy and system delay. Ultimately, the simulation results substantiate the efficacy of the proposed algorithm.

\begin{figure*}[ht]
\centering
\includegraphics[width=6in]{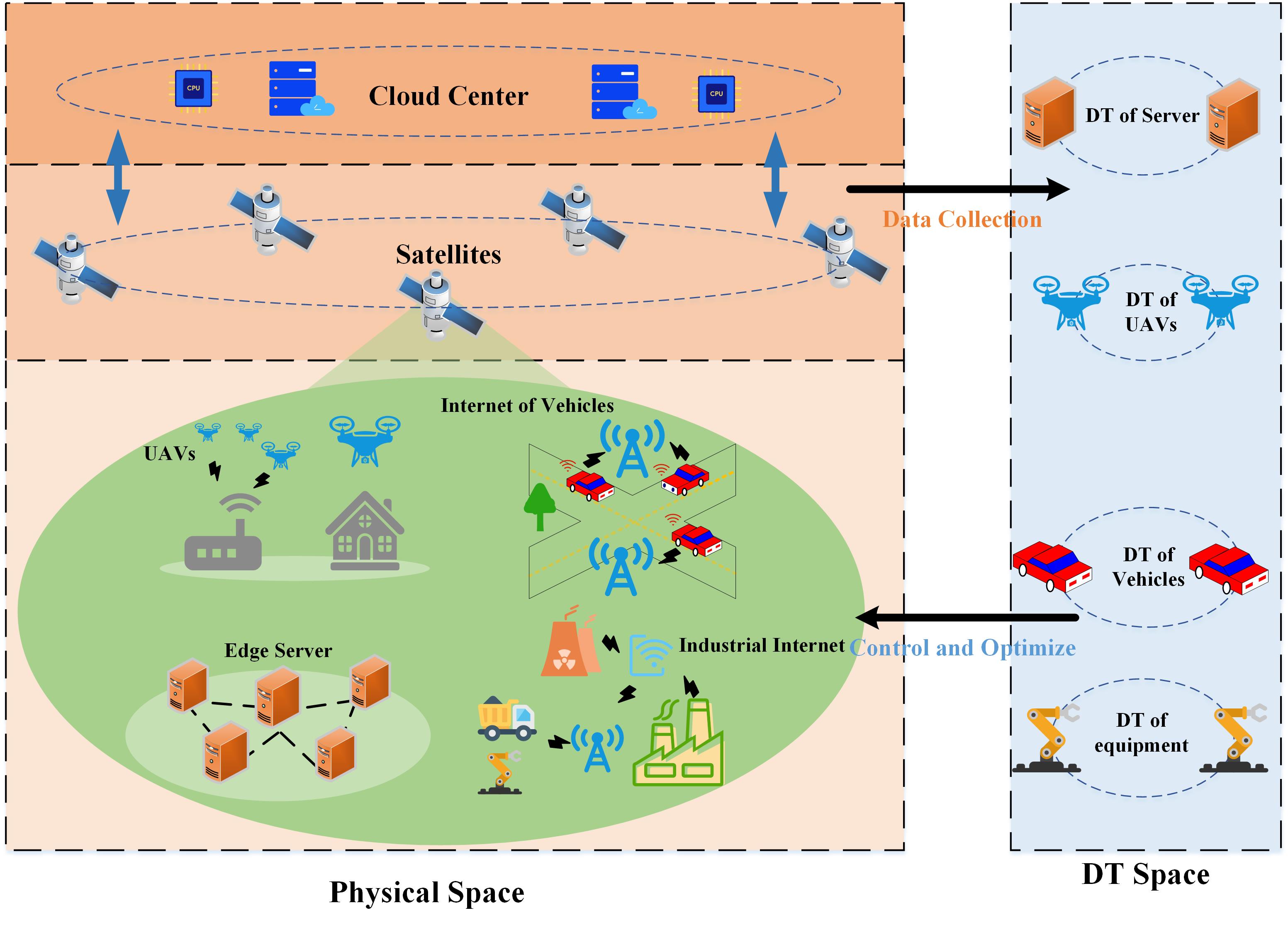}
\caption{Digital twin network architecture of satellite-terrestrial integrated network.}
\label{fig_1}
\end{figure*}

\begin{figure}[tbh]
	\centering
	\includegraphics[scale=0.45]{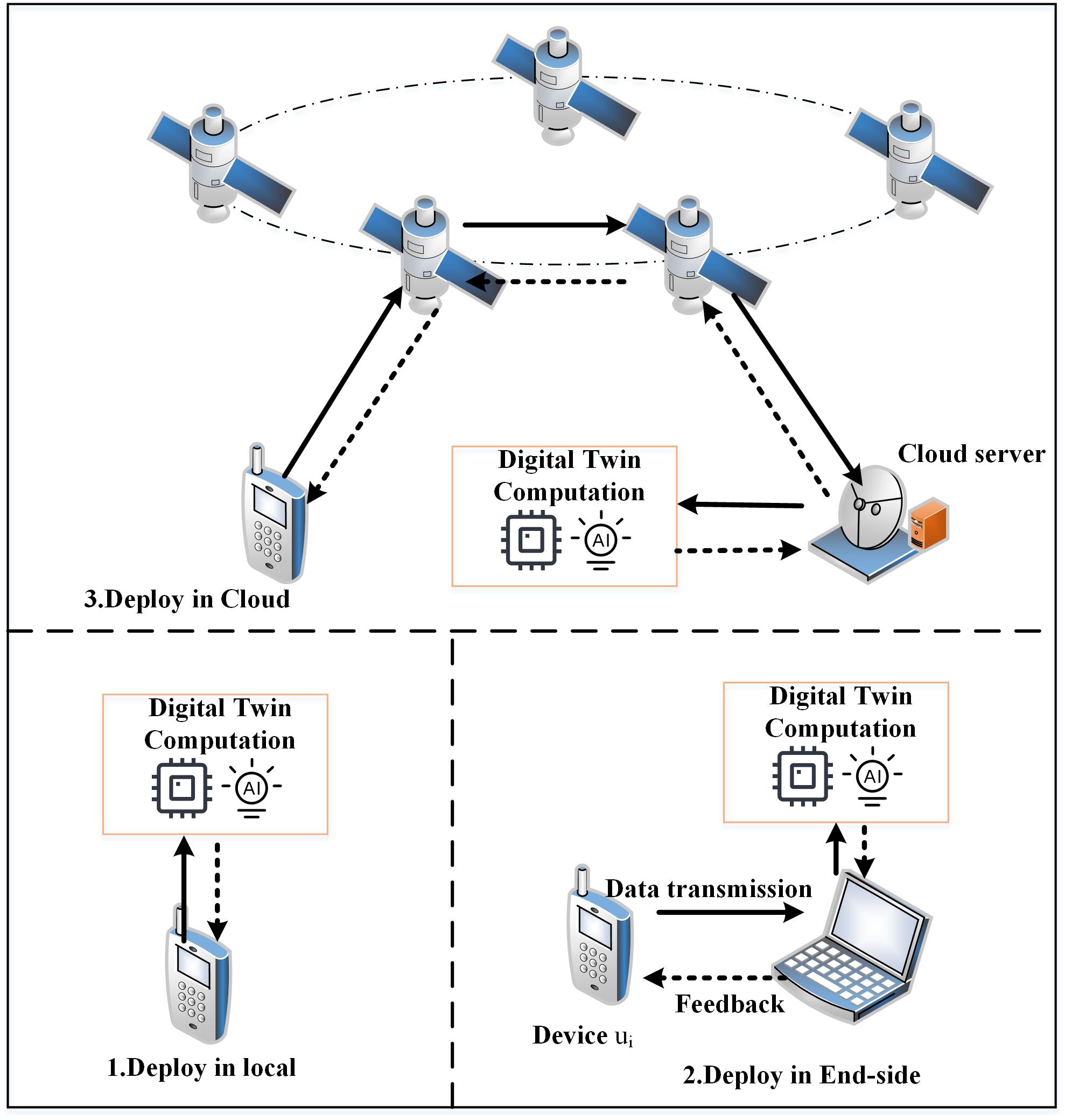}
	\caption{Process of constructing a digital twin } \label{fig:Prediction}
\end{figure}

\section{Related work}
DTs can virtually model physical entities, and deploying DTs in the network can then monitor, analyze, predict, and optimize the network. \cite{22} proposed the use of DTs to optimize user association, resource allocation, and offloading probability in MEC to reduce system energy consumption. \cite{23} integrated artificial intelligence and DT technology principles into the architecture of a  edge computing network, showcasing an innovative approach that combines these advanced technologies for enhanced performance and functionality, and focus on matching the potential edge services. \cite{24} introduced a strategy where mobile users can intelligently delegate computational tasks to servers, facilitated by the utilization of DTs. \cite{25} devised a novel unmanned aerial vehicle-augmented mobile network infrastructure, tailored to ensure seamless and efficient communication services for all mobile users operating within congested, high-density traffic environments, and introduced an online training-based DT authorization mechanism for dynamic resource allocation. For the relevant satellite network schemes, to realize an intelligent sky and ground integrated vehicle network, \cite{42} designed an architecture that supports DT technology. \cite{43} used DT technology to create a virtual network for a physical satellite network, optimize network performance, and solve the limitation of low reliability of traditional satellite simulation platform. \cite{44} supported the optimal multi-beam design of a 6G satellite-ground integrated network according to the requirements of critical tasks and combined with DT technology. \cite{45} presented the challenges of meeting the requirements of DTs in the STIN, and analyzed the technology of implementing DT system. Our proposal adopts the DT multi-deployment learning method.

For the problem of deploying DTs in the network to improve the performance of a physical system, \cite{26} proposed a wireless DITEN to effectively build and maintain DTs in a wireless digital twin network, the DT placement problem is proposed for dynamic network status and changing network topology. \cite{27} designed a dynamic edge network DT model that captures the dynamic and service requirements of a real-time edge network and realizes the efficient deployment of DT servers. \cite{28} considered the deployment details of the edge network, device types and their social characteristics, and solves the dynamic layout of DTs with social ability at the edge. To enhance the property of the DT model, \cite{29} proposed a DT architecture with wireless computing capability, and the placement and migration of DTs are solved by considering synchronization delay and DT error. \cite{30} considered the placement of DTs to meet the requirements to provide response delay minimization to all users. 

The above studies provide reference solutions for the placement and deployment of DTs. Distinct from the aforementioned research efforts, this study considers how to deploy DTs in a STIN. To optimize the utilization of available resources in the STIN, we consider a multi-layer deployment method. Our goal is to minimize the delay between users and DTs.

\section{System model}

Firstly, We introduce the architecture of the DT-empowered STIN model, and then study the communication and computing models in the system to formulate the DT multi-layer deployment problem.

\subsection{Architecture of DT-empowered satellite-terrestrial integrated network system}

We propose an architecture that introduces DT technology in a STIN, as shown in Fig. 1. The architecture includes two spaces: a physical space and DT space. With the help of DT nodes, terminal devices, along with satellites and cloud servers, comprise the diverse entities of the physical layer, which can be systematically mapped and modeled to generate DT models. This process enables the creation of DT models that accurately mirror the composition and operations of these physical devices. Satellites can provide seamless coverage of network services for terminal devices, and cloud servers have relatively sufficient computing resources. DT nodes are selected from nodes with computing power for deploying DT models.

In the STIN, the association between terminal devices and DT nodes can be established to meet the needs of communication and computing resources. DTs realize accurate mapping of physical entities through data transmission, DT modeling, and feedback optimization. Through DTs, the historical state information and real-time operation status of terminal devices can be collected to monitor, predict, and analyze them. Each DT model in the DT layer can map the real-time operation status of terminal devices. In addition, we assume that the DT corresponding to terminal devices is deployed on resource-rich computing nodes. In this architecture, we select an optimized computing node as the deployment location of the DT model for different terminal devices to build the DT model. Let I = $\left \{1, \ldots, i, \ldots, I  \right \}$  and J = $\left \{1, \ldots, j, \ldots, J  \right \}$ to represent the terminal devices that are needed to build a DT and the nodes with computing resources, there is some overlap between them. This is because some terminal devices need the DT services and have free computing resources to provide to other terminal devices to maximize the utilization of computing resources.

The DT multi-layer deployment problem is established according to the DT model construction method in this study. DT deployment is the first step in constructing a DT, and includes finding the best location for deploying the DT model for each terminal device from multi-layer computing nodes, to ensure low-latency interaction between physical entities and the DT model.

\subsection{Communication and computation model}
Fig. 2 shows the complete procedure of deploying the user i DT in different nodes. In our system, the DT can be deployed locally, by other terminal devices, or by satellite transmit to the cloud server (CS). The delay in deploying a DT mainly includes three issues, transmission delay, computation delay, and feedback delay. The data of user i is transferred to the corresponding compute node responsible for deploying the DT, and then the compute node uses the computing power to efficiently process and interpret the acquired data, subsequently constructing a personalized DT model  specifically for user i, as shown in Fig. 2. Ultimately, the feedback results are transmitted back to the user; the feedback results meet the services requirements of the user or have certain optimization effects. The specific communication and calculation models are as follows:

\noindent \textbf{Deploy in local:} If the local terminal has certain computing resources, the DT can be deployed on the local terminal without communication delay. The delay calculation is

\begin{equation}
{{L}_{\text{i}}}=\frac{\mu {{D}_{\text{i}}}}{{{\text{f}}_{\text{i}}}}.
\end{equation}

where ${{D}_{\text{i}}}$ is the data size of the DT of user i and ${{\text{f}}_{\begin{smallmatrix} \text{i} \\ \end{smallmatrix}}}$ is the computing resource of user i. $\mu$ is the computing resource required for processing each bit of data.

\noindent \textbf{Deploy in end-side:} The DT can be deployed on other terminal devices with rich computing resources. Communication between terminal devices are via Bluetooth or Wi-Fi. The communication between the terminal and the end-side includes uplink communication for transmitting data to the DT node and downlink communication for transmitting feedback results from the end-side node back to the user. The feedback results size returned to the user is much smaller than the DT data that the user needs to transmit, so only the uplink communication delay is considered in the communication model \cite{31}. Wireless communication is adopted between the user and the end-side devices, and the maximum theoretical communication transmission rate is
\begin{equation}
{{\text{r}}_{\text{ij}}}=W\text{log}(1+\frac{{{P}_{\text{i}}}{{G}_{\text{ij}}}}{W{{N}_{0}}}).
\end{equation}
where ${{P}_{\text{i}}}$ represents the transmit power, ${{G}_{\text{ij}}}$ represents the channel power gain between user i associating with the computing node j, W is the channel bandwidth, ${{N}_{0}}$ is the noise power spectral density, and the transmission delay from user i to the end-side devices is

\begin{equation}
L_{\text{ij}}^{\text{com}}=\frac{{{D}_{\text{i}}}}{{{\text{r}}_{\text{ij}}}}.
\end{equation}

The total computing resources of end-devices j are expressed as ${{F}_{\text{j}}}$. You can allocate computing resources from end-devices with sufficient computing resources to multiple users to maintain their DT deployed on end-devices j. Set computing resources ${{\text{f}}_{\text{ij}}}$, which represent the DT assigned to user i. The time required for user i to process the DT data can then be expressed as

\begin{equation}
L_{\text{ij}}^{\text{cmp}}=\frac{\mu {{D}_{\text{i}}}}{{{\text{f}}_{\text{ij}}}}.
\end{equation}

where $\sum\limits_{\text{i}=1}^{N}{{{\lambda }_{\text{ij}}}{{\text{f}}_{\text{ij}}}\le {{F}_{\text{i}}}}$, if ${{\text{f}}_{\text{ij}}}>0$, ${{\lambda }_{\text{ij}}}=1$ , otherwise ${{\lambda }_{\text{ij}}}=0$. Based on the preceding formula, the delay for deploying the DT at the end is 

\begin{equation}
{{L}_{\text{ij}}} = \frac{{{D}_{i}}}{{{r}_{\text{ij}}}}+\frac{\mu {{D}_{i}}}{{{f}_{\text{ij}}}}.
\end{equation}

\noindent \textbf{Deploy in cloud:} DTs can be deployed on the cloud server (CS). When users cannot connect to the CS through the ground network, DT data can be transmitted to the CS through the STIN by the backhaul link. In our system model, user i can connect to the satellite directly through wireless communication, and then the satellite will return to the ground gateway station, which is connected to the CS. DT data processing and modeling is performed in the CS. The computing resources in the CS are sufficient, but there is a large communication delay.

For wireless communications between users and satellites, the channel capacity of the connection between them is limited by the distance between them, transmission power, signal-to-noise ratio and other parameters. The available spectrum for wireless data transmission from the user to the covered satellite is equally distributed among all users, expressed as ${{W}^{\text{is}}}$. Our framework employs a quasi-static approach, where the channel conditions and network topology are assumed to be stable within a given time slot, facilitating the analysis and modeling process \cite{41}, ${{G}_{\text{is}}}$ represents the channel gain, using $P_is$ to represent the transmission power, the uplink data rate from user i to satellites is

\begin{equation}
{{\text{r}}_{\text{is}}}={{W}^{\text{is}}}\text{log}(1+\frac{{{P}_{\text{is}}}{{G}_{\text{is}}}}{{{W}^{\text{su}}}{{N}_{0}}}).
\end{equation}

After the satellite receives the user's data, it transmits the data to the CS through the ground gateway station. Similar to Eq.(5), the data transmission rate from the satellite to the ground gateway station is

\begin{equation}
{{\text{r}}_{\text{sc}}}={{W}^{\text{sc}}}\text{log}(1+\frac{{{P}_{\text{sc}}}{{G}_{\text{sc}}}}{{{W}^{\text{sd}}N_{0}}}).
\end{equation}

where ${{W}^{\text{sc}}}$ represents the bandwidth assigned to the satellite by the ground gateway station, ${{P}_{\text{sc}}}$ is the transmitting power of the satellite, and ${{G}_{\text{sc}}}$ is the channel gain between them. The transmission delay for transferring $D_{i}$ is

\begin{equation}
L_{\text{ic}}^{\text{com}}=\frac{{{D}_{i}}}{{{r}_{\text{is}}}}+\frac{{{D}_{\text{i}}}}{{{\text{r}}_{\text{sc}}}}+\frac{{{\text{d}}_{\text{is}}}+{{\text{d}}_{\text{sc}}}}{\text{c}}.
\end{equation}

where $c$ is the speed of light, and ${{\text{d}}_{\text{is}}}$ and ${{\text{d}}_{\text{sc}}}$ are the distance between user i and the satellite and the distance between the satellite and the CS, respectively. The calculation delay in the CS is expressed as 

\begin{equation}
L_{\text{ic}}^{\text{cmp}}=\frac{\mu {{D}_{i}}}{{{f}_{\text{ic}}}}.
\end{equation}

The total delay for user i is expressed as

\begin{equation}
{{L}_{\text{ic}}}=L_{\text{ic}}^{\text{com}}+L_{\text{ic}}^{\text{cmp}}.
\end{equation}

\begin{table}[thb]
\scriptsize
    \renewcommand{\arraystretch}{1.2}
    \setlength{\tabcolsep}{2pt}
    \caption{Summary of main notations}
    \label{table_example}
    \centering
    \begin{tabular}{ll}
        \hline
        Notation & Description  \\
        \hline
        I & The set of users   \\
        J & The set of computing nodes   \\
        $L_i$ & The latency of user i deploy in local \\
        $\mu$ & The computing resource for processing one bit data \\
        $D_i$ & The data size of user i\\
        $f_i$ & The computing resource of user i \\
        $r_{ij}$ & The achievable data rate between user i and node j \\
        W & The bandwidth of the transmission channel \\
        $P_{i}$ & The transmission power of user i \\
        $G_{ij}$ & The channel power gain between user i and node j \\
        $L_{ij}^{com}$ & The communication latency between user i and node j \\
        $L_{ij}^{cmp}$ & The computation latency of tasks for user i in node j \\
        $r_{is}$ & The achievable data rate between user and satellite \\
        $W^{is}$ & The transmission bandwidth between user i and satellite\\
        $P_{is}$ & The transmission power of user i to satellite \\
        $G_{is}$ & The channel power gain between user i and satellite \\
        $r_{sc}$ & The achievable data rate between satellite and cloud \\
        $W^{sc}$ & The transmission bandwidth between satellite and cloud \\
        $P_{sc}$ & The transmission power of satellite to cloud \\
        $G_{sc}$ & The channel power gain between satellite and cloud \\
        $L_{ic}$ & The communication latency between user i and cloud \\
        $S^t$ & The MARL state at time slot t \\
        $A^t$ & The MARL action at time slot t \\
        $R_t$ & The MARL reward at time slot t \\
        \hline
    \end{tabular}
\end{table}

\section{Problem formulation}
We formulate the DT multi-layer deployment problem by considering the different characteristics of compute nodes located at different layers. DT deployment needs to consider the delay caused by data transmission and the establishment of DT models. In our system model, the dynamic network state changes are considered, and the optimal DT deployment location is selected for each user to maximize the system performance.

The constraints of the optimization problem are described below. We use $G(U,N,DT,C)$ to represent our satellite-terrestrial integrated DT network, where $U$ is the user who needs to build the DT, N is the node with computing power, $DT$ is the node where the DT is deployed, and $C$ is all the communication links in the network. Capacity $N_j$ of the DT node indicates the maximum number of DTs that the node can maintain.
We use the matrix $ \lambda=$

\begin{equation}
\left[\begin{array}{llll}
\lambda_{11} & \lambda_{12} & \cdots & \lambda_{1 J} \\
\lambda_{21} & \lambda_{22} & \cdots & \lambda_{2 J} \\
\cdots & \cdots & \cdots & \cdots \\
\lambda_{I 1} & \lambda_{I 2} & \cdots & \lambda_{I J}
\end{array}\right]
\end{equation}

to represent the correspondence between the user and the deployed DT, where $\lambda_{ij} = 1$ if the DT of user i is deployed on node j, otherwise $\lambda_{ij} =0$. The total system delay is an indicator to measure the DT deployment policy and represents the operating performance of the system. The total deployment delay is

\begin{equation}
{{L}_{\text{sum}}}=\sum\limits_{\text{i}\in \text{I,j}\in \text{J}}{{{\lambda }_{\text{ij}}}}{{L}_{\text{ij}}}.
\end{equation}

In the above formula, the system delay is determined by the DT deployment strategy and the deployment delay between users and DT nodes, and the deployment delay is affected by compute nodes located at different layers.
The goal of the DT multi-layer deployment problem is to find the right deployment location to minimize system latency, and deployment at different layers directly affects system latency performance. Therefore, the optimization problem is expressed as follows:

\begin{subequations}\label{eq:ctr_shale}
\begin{align}
&\text{min}\quad{{L}_{\text{sum}}}. \\
\text{s.t.} & {\lambda }_{\text{ij}}\in \{0,1\},  \\
    &\sum\limits_{\text{j}\in J}{{{\lambda }_{\text{ij}}}}=1,\forall \text{i}\in I,  \\
    &\sum\limits_{\text{i}\in I}{{{\lambda }_{\text{ij}}}}\le {{N}_{\text{j}}},\forall\text{j}\in J, \\
    &0 \leq \sum_{\mathrm{i} \in I} \lambda_{\mathrm{ij}} \mathrm{f}_{\mathrm{i}} \leq \mathrm{f}_{\mathrm{j}}^{\max }, \forall \mathrm{j} \in J.
\end{align}
\end{subequations}

Constraint (13b) indicates that variable ${{\lambda }_{\text{ij}}}$ has only two states, 0 and 1, which are undeployed and deployed, respectively. Constraint (13c) indicates that a user's DT can be deployed on only one DT compute node. Constraint(13d) indicates that the number of DTs deployed on compute nodes cannot exceed the capacity. Constraint (13e) means that the computing capacity allocated to each compute node cannot exceed its maximum computing capacity. To obtain a satisfactory deployment strategy, it is necessary to search in a space consisting of $J^I$ possible deployment decisions. Traditional optimization methods, such as game theory and convex optimization, have difficulty in guaranteeing the long-term performance of the resulting decisions, have high complexity, and are not suitable for the network with rapidly changing environments considered in this study. To address the problem posed by of Eq. (13a), it is essential to find the optimal decision in each time slot. The model-free reinforcement learning method adopted in this study is a very promising approach for learning the optimal strategy through direct interaction with dynamic and decentralized environments when dealing with tasks with unknown environments. Specifically, through an algorithm based on MARL, for constraint (13c), each agent will only select the action space where the only one action is 1. For constraint (13d), when the number of DTs deployed in this compute node j reaches the maximum, the agent will not choose to continue to deploy on this node. For constraint (13e), computing resources in a compute node are evenly distributed among all agents.

\section{MARL-based algorithm for the multi-layer deployment problem in a satellite-terrestrial integrated network}
We propose a MARL-based algorithm to solve the proposed digital twin multi-layer deployment problem. The MARL framework will be introduced in the STIN considered in this study to maximize system performance. We treat the formulaic problem in the above equation as a distributed MARL task and set each user as one agent. This is explained in detail below.

\begin{figure}[tbh]
	\centering
	\includegraphics[scale=0.7]{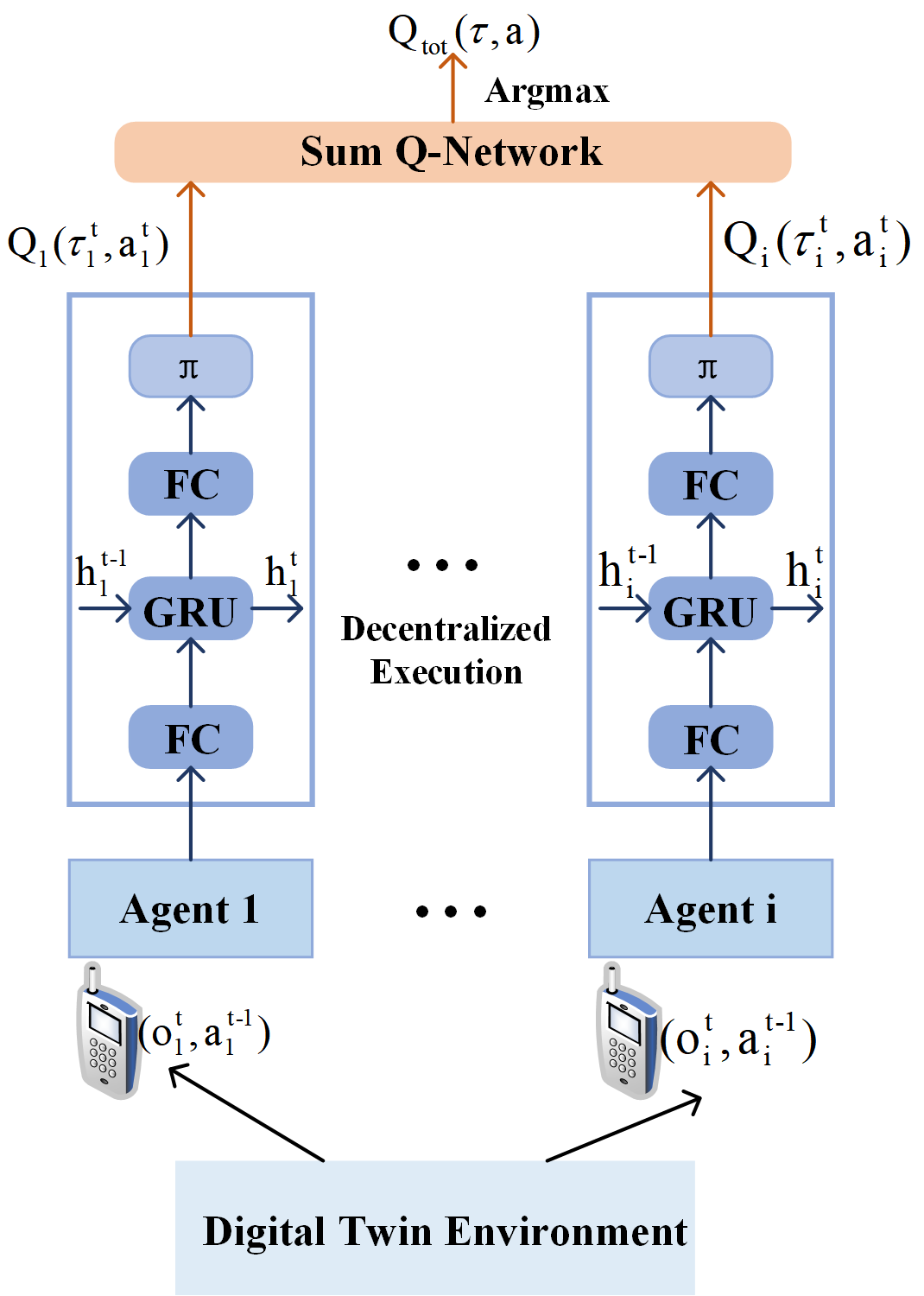}
	\caption{The MARL method for the DT deployment structure. } \label{fig:Prediction}
\end{figure}

\subsection{Problem transformation}
Reinforcement learning does not require datasets, but rather learns through interaction with the environment. It follows a trial-and-error approach, optimizing its decision-making strategies by constantly exploring the environment and learning from experience. In this kind of communication network, with the increase of users number, the application of traditional single-agent reinforcement learning methodologies in complex and ever-evolving environments poses a significant challenge, as it may cause the agent to unduly focus its attention solely on the immediate behavior of its competitors, while ignoring other potential behavior patterns or strategies. The MARL algorithm in this paper can help the agent to consider the existence and behavior of other agents in the learning process, make more intelligent decisions, and achieve better system performance.

The issue presented in Eq. (13a) can be reframed as a decentralized partially observable Markov decision process (Dec-POMDP) challenge specifically tailored for DT users,  each corresponding to an agent, each of which can only observe a local space at each time slot \cite{33}. Each agent independently selects actions based on observed local spatial states to maximize system rewards. This problem can be formulated as $\langle S, \left\{O_{\mathrm{i}}\right\}_{\mathrm{i} \in I}, \left\{A_{\mathrm{i}}\right\}_{\mathrm{i} \in I} , R , \gamma\rangle$ , where $S$ is the system state, $O_i$ is the state space observed locally by agent i, $A_i$ is the action space of agent i, the reward function is denoted as R, and the discount factor is denoted as $\gamma$. Each agent first performs an action $\text{a}_{\text{i}}^{\text{t}}\in {{A}_{\text{i}}}$ based on the locally observed state $\text{o}_{\text{i}}^{\text{t}}\in {{O}_{\text{i}}}$, and then the joint action can be obtained, expressed as ${{\text{a}}^{\text{t}}}\in A={{A}_{1}}\times \ldots \ldots \times {{A}_{\text{i}}}$. Following the execution of a collective action at time slot t, the environment computes a global reward ${{\text{r}}^{\text{t}}}=R({{\text{s}}^{\text{t}}},{{\text{a}}^{\text{t}}})$ and proceeds to update the system state to ${{\text{s}}^{\text{t}+1}}$. In the subsequent sections, we introduce the constructs of the system state space S, the individual agent's observation spaces ${{\{{{O}_{\text{i}}}\}}_{\text{i}\in I}}$, the action spaces ${{\{{{A}_{\text{i}}}\}}_{\text{i}\in I}}$ each tailored to a user, and the reward function R that evaluates the effectiveness of actions.

(1) System state: In our proposed scheme, the system state includes the physical state of all users ${{\varphi }^{\text{t}}}$, the network resources state ${{\beta }^{\text{t}}}$, and deployment policy decisions ${{\lambda }^{\text{t}}}$. The physical status of the user includes the distance, the data to be transferred of user i, and the data transmission rate. We represent the system state as

\begin{equation}
{{\text{S}}^{\text{t}}}=\{{{\varphi }^{\text{t}}},{{\beta }^{\text{t}}},{{\lambda }^{\text{t}}}\}.
\end{equation}

(2) Action space: Agents are users in our system. Each agent selects its action, including the DT deployment policy. The deployment policy of user i is $\left \{ \lambda_{i1}, \lambda_{i2}, \ldots, \lambda_{ij}\right \}$ , where $\lambda_{ij}$ = 1 if the DT of end user i is deployed to compute node j. Else, $\lambda_{ij}$ = 0. The action spaces across all multi agents can be collectively represented as

\begin{equation}
{{A}^{\text{t}}}=\{{{\text{a}}_{\text{i}}}|{{\text{a}}_{\text{ij}}}={{\lambda }_{\text{ij}}},\forall \text{i}\in I,\forall \text{j}\in J\}.
\end{equation}

(3)Reward function: In this MARL, all agents share the reward. When a joint action is taken, the environment returns a reward. The primary objective for each agent is to expedite the deployment of its DT by minimizing associated delays. The overall reward function encompasses two key components: a delay-based reward function $R_L$ and a deployment cost function $R_C$. Here, we elaborate the delay-based reward function is

\begin{equation}
{{R}_{L}}=-{{L}_{\text{sum}}}.
\end{equation}

The cost incentive function for deploying DT services is

\begin{equation}
{{R}_{C}}=\phi \cdot \text{m}.
\end{equation}

where $\phi$ is defined as unit expense coefficient for deploying the DT services, and m is the number of DTs. The total reward function is

\begin{equation}
{{R}_{\text{t}}}=\alpha \cdot {{R}_{L}}-\beta \cdot {{R}_{C}}.
\end{equation}

where $\alpha$ and $\beta$ acting as the respective balancing coefficients for delay and cost rewards, we introduce the concept of applying MARL as a means to optimize the search for the most effective DT deployment strategy, where all agents collaborate to make deployment decisions. The goal is to maximize the global system reward, which is described as follows

\begin{equation}
R=\sum\limits_{\text{t}=1}^{T}{{{\gamma }^{\text{t}-1}}{{R}_{\text{t}}}}.
\end{equation}

where $\gamma \in(0,1]$ signifies the reward discount factor, and each agent i acts according to the local observation state and its local strategy in the partial observation environment. We use a federation strategy $\pi =[{{\pi }_{1}}, {{\pi }_{2}}, \ldots, {{\pi }_{\text{i}}}]$ that represents i agents. The joint action function is

\begin{equation}
{{Q}^{\pi }}({{\text{s}}^{\text{t}}},{{\text{a}}^{\text{t}}})=E[{{R}_{\text{t}}}|{{\text{s}}^{\text{t}}},{{\text{a}}^{\text{t}}}].
\end{equation}

where $E[\cdot ]$ signifies the expected value operation, whereas the joint action function, denoted as ${{Q}^{\pi }}({{\text{s}}^{\text{t}}},{{\text{a}}^{\text{t}}})$, quantifies the anticipated global reward when commencing from $s^t$ and $a^t$ is followed by the strategy $\pi$, and the optimal union when ${{Q}^{\pi }}({{\text{s}}^{\text{t}}},{{\text{a}}^{\text{t}}})$ is the largest is slightly $\pi^*$.

\subsection{MARL for DT deployment}
The agent in a traditional centralized reinforcement learning algorithm updates its policy independently as learning progresses, which makes the convergence of the algorithm difficult to \cite{33}. Therefore, we propose to use a MARL-based approach to solve our DT deployment problem.

This algorithm adopts a centralized training decentralized execution approach, which can use the system's global environment information to train agents centrally, and each agent chooses to deploy actions in a decentralized manner according to the local observation environment \cite{35}. However, the local observed state $(o_i,a_i)$ cannot describe the environment state, and historical observations can be stored to help the function Q. The historical observation results of agent joint action are expressed as ${{\tau }_{\text{t}}}=(\tau _{\text{t}}^{1}, \tau _{\text{t}}^{2}, \ldots, \tau _{\text{t}}^{I})$. The joint actions of all agents are represented as $a_{t}=(a^{1}_t,a^{2}_t, \ldots, a^{I}_t)$. Hence, the primary objective of this algorithmic approach is to acquire the cumulative action-value function ${{Q}_{\text{sum}}}({{\tau }_{\text{t}}}, {{\text{a}}_{\text{t}}})$ that integrates global information during the centralized training phase, enabling a comprehensive evaluation of joint actions. By combining action value functions, distributed agents can choose actions that are closer to the global optimal solution. The architecture of the presented algorithm, as depicted in Fig. 3, comprises the agent network module and the integrated hybrid network module.

\begin{algorithm}[t]
    \caption{The MARL-based DT deployment algorithm}
    \label{alg:AOA}
    \renewcommand{\algorithmicrequire}{\textbf{Input:}}
    \renewcommand{\algorithmicensure}{\textbf{Output:}}
    \begin{algorithmic}[1]
        \REQUIRE The user set I,the compute node J  
        \ENSURE The DT server deployment strategy    

        \FORALL{agents $i \in I$}
            \STATE Initialize network parameter and replay buffer
        \ENDFOR

        \FOR{episode $e = 1 \text{ to } E$}
            \STATE Initialize system state $S^t$
            \FOR{time step t}
               \FORALL{agents $i \in I$}
                  \STATE Determine actions $a_i^t = max_aQ(o_t,a;\theta_i)$ based on the observation $o^t$
                \ENDFOR  
            \STATE Execute joint action $\mathbf{a}^t = (a_1^t, \ldots, a_I^t)$ and receive joint reward $r_t$ and new observations $\mathbf{o}^{t+1}$
            \FORALL{agents $i \in I$}  
                \STATE Store transition $(o_i^t, a_i^t, r_t, o_i^{t+1})$ in $\mathcal{D}_i$  
            \ENDFOR 
            \FORALL{agents $i \in I$}  
                \STATE Update $\theta_i$ using gradient descent based on transitions sampled from $\mathcal{D}_i$  
                \STATE Update $\theta_i^-$ towards $\theta_i$  
            \ENDFOR 
            \ENDFOR
        \RETURN The optimized DT server deployment strategy based on learned policies
        \ENDFOR
    \end{algorithmic}
\end{algorithm}

(1) Agent Q network: Each agent incorporates a gate recurrent unit (GRU) alongside fully connected (FC) layers. The agent processes its local observation $o^{t}_i$ and the previous action $a_{i}^{t-1}$
as inputs, subsequently generating the local action-value function ${{Q}_{i}}(\tau _{i}^{t},a_{i}^{t})$ as its output.In the Q network of each agent, $h_i^{t-1}$ plays a crucial role. As the output of the GRU, $h_i^{t-1}$ not only contains the internal state information of agent i at time step t-1, but also integrates all the historical observation and action information from the initial time step to t-1. This ability allows the agent to learn and make decisions effectively in a partially observable environment, even when the observed value $o^{t}_i$ of the current time step may not be sufficient to fully reveal the state of the environment. In a partially observable network, historical observations can be stored in a network, which contributes to the agent's ability to learn over longer timescales \cite{34}.

(2)Mixing network: The MARL algorithm based on value decomposition dissects the cumulative action-value function ${{Q}_{\text{tot}}}({{\tau }_{\text{t}}},{{\text{a}}_{\text{t}}})$ into individual local action-value functions ${{Q}_{\text{i}}}(\tau _{i}^{t},a_{i}^{t})$ to guide the strategy of each agent separately \cite{32}. The joint action value function is defined as follows

\begin{equation}
{{Q}_{\text{tot}}}({{\tau }_{\text{t}}},{{\text{a}}_{\text{t}}})\approx \sum\limits_{\text{i}\in I}{{{Q}_{\text{i}}}(\tau _{i}^{t},a_{i}^{t};\theta _{i}^{\text{t}})}.
\end{equation}

where $\theta _{\text{i}}^{\text{t}}$ are the Q network parameters. The constraints between $Q_{tot}$ and any single action value $Q_i$ need to be satisfied as follows

\begin{equation}
\frac{\partial {{Q}_{\text{tot}}}}{\partial {{Q}_{\text{i}}}}\ge 0,\forall \text{i}\in A.
\end{equation}

we express the target Q value as $Q_{\text{i}}^{-}$, thus $Q_{\text{tot}}^{-}({{\tau }_{\text{t}}},{{\text{a}}_{\text{t}}})\approx \sum\limits_{\text{i}\in I}{Q_{\text{i}}^{-}(\tau _{i}^{t},a_{i}^{t};\theta _{i}^{\text{t}})}$. The time difference target is represented by ${{\text{y}}_{\text{tot}}}=\text{r}+\gamma \text{ma}{{\text{x}}_{{{\text{a}}^{'}}}}Q_{\text{tot}}^{{}}({{\tau }^{'}},{{\text{a}}^{\text{ }\!\!'\!\!\text{ }}};{{\theta }^{-}})$, where ${{\theta }^{-}}$ is the target network parameter.
The overall loss function is

\begin{equation}
L(\theta )=\frac{1}{X}\sum\limits_{\text{x}=1}^{X}{{{(\text{y}_{tot}^{x}-Q_{tot}^{x}({{\tau }^{x}},{{a}^{\text{x}}}))}^{2}}}.
\end{equation}

where $X$ denotes the quantity of small batch sizes that are randomly drawn from the replay buffer, and $x$ signifies the specific instance within this collection, representing the x-th sampled batch. The pseudo-code for the MARL-based DT deployment algorithm is shown in Algorithm 1.

\subsection{Computational Complexity Analysis}

The computational complexity (CC) of the proposed DT deployment algorithm is analyzed. First, the initialized network parameters are input into the agent network to generate the initial network environment, which requires O(1). The centralized hybrid training process can be implemented in the main server with sufficient computing power. Thus, we mainly consider the CC of the algorithmic distributed process. For the multi-agent distributed execution process, each action and local observation are evaluated by the neural network to generate the corresponding local action value function Q. We employ $N_1$, $N_2$, $N_3$, and $N_4$ to respectively represent the number of neurons per neural layer in the neural network, and the number of hidden neural layers is represented by U. Hence, the computational complexity for each agent is O($N_1N_2 + (U - 1)N_2N_3 + N_3N_4$). For each neuron, an activation function needs to be applied before the output. Therefore, the total computational cost of the activation function can be approximated as O($N_2U$), because each neuron must perform an activation function calculation. By summing up the computational cost of matrix multiplication and the computational cost of the activation function, the total CC of the neural network is O($N_1N_2 + (U - 1)N_2N_3 + N_3N_4 + N_2U$). Based on the aforementioned analysis, the CC of our algorithm can be approximated as O($N_2(N_1 + UN_3) + N_3N_4$).

\section{Simulation results}
This section provides several numerical simulations to verify the performance of the MARL-based DT deployment scenario. We compare this algorithm to the following three benchmarks.

(1) Independent Q-Learning(IQL) scheme: In the IQL algorithm, each agent trains and makes decisions based on local observations \cite{37}.

(2)QMIX scheme: In the QMIX algorithm, the individual action value functions are summed to derive the combined joint action value function by using hypernetwork weighted summation \cite{35}.

(3)RD scheme: In this algorithm, each user randomly selects the deployment node.

\subsection{Simulation settings}

\begin{table}[thp]\footnotesize
\centering
\caption{Simulation parameters} \label{Table:2}

\begin{tabular}{|l|l|}
\hline
\textbf{Parameter}        & \textbf{Value}       \\ \hline
Number of users           & 20                   \\ \hline
Number of end-side devices & 3                    \\ \hline
Data size of users $D_i$        & {[}0.5, 2{]} MB      \\ \hline
Workload of one-bit data $\mu$  & {[}50, 150{]} cycle/bit      \\ \hline
Computation capacity $f_i$      & {[}0.5, 20{]} GHz \\ \hline
Noise power function $N_0$      & - 174 dBm/Hz           \\ \hline
Training iterations       & 1,000                 \\ \hline
Discounting factor $\gamma$        & 0.9                  \\ \hline
Learning rate             & 0.0001               \\ \hline
Batch size                & 64                   \\ \hline
\end{tabular}
\end{table}

We consider simulating a network scenario with 20 users, 3 end-side devices, and 1 satellite that can backhaul user data to a ground cloud server. Users and terminal devices are randomly distributed in the 500 m × 500 m area. In our simulation, user status and network channel status change over time. The maximum transmission power of the user devices is 200mW, the training rounds in our method is 1000, and the discount factor $\gamma$ is set to 0.9. We set the weight of the reward function $\alpha$ and $\beta$ to 0.7/0.3, the weight of the reward function will affect the decision of the agent. This study focuses on reducing the system delay. A comprehensive listing of the simulation parameters, can be found in Table 2 for detailed reference.

\subsection{Results and analysis}

Fig. 4 shows the convergence curves of the three algorithms as the training rounds increases. The proposed algorithm obtains the highest reward value, followed by the IQL and QMIX algorithms. The complexity of the IQL and QMIX algorithms is approximately equal to the algorithm in this study. Most operations involve matrix multiplication and addition, and the specific algorithm complexity is related to the structure of the neural network in the agent network \cite{36,37}. The reason is that agents in the IQL scheme explore and learn independently in a shared environment, as a result, agents lack the ability to discern whether alterations in the system's environment stem from their individual actions or are a consequence of the behaviors executed by other agents within the system. This strategy of multiple agents learning independently in a non-stationary environment will lead to a decline in decision-making quality. Therefore, the efficiency of this algorithm is inferior to the proposed algorithm. The hybrid network can learn the weight of the local action value function from the global action value. In our system, the Q value weights of each agent are similar; therefore, the QMIX algorithm is not suitable for our system and the reward value is the lowest.

\begin{figure}[htb]
	\centering
	\includegraphics[scale=0.36]{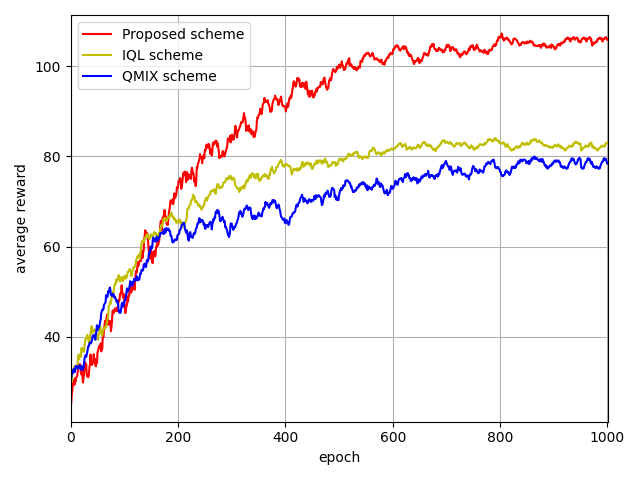}
	\caption{Average reward.} \label{fig:Prediction}
\end{figure}

\begin{figure}[htb]
	\centering
	\includegraphics[scale=0.36]{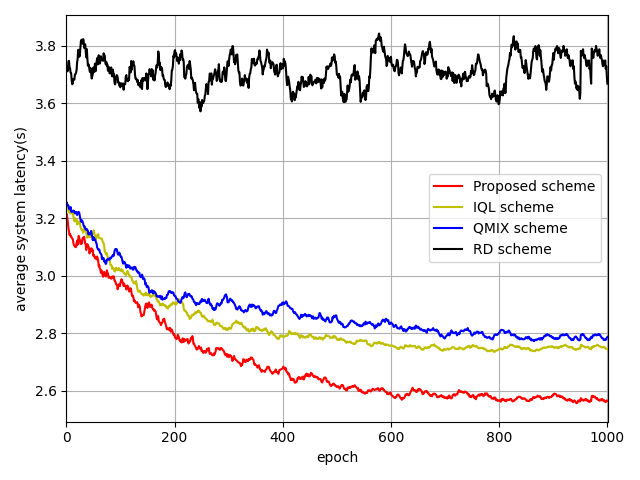}
	\caption{Epochs vs average system latency for different algorithms.} \label{fig:Prediction}
\end{figure}

Fig. 5 shows that when different schemes are adopted, the average system delay changes with increased training rounds. Fig. 5 shows that the average system delay is the lowest when the proposed scheme is adopted. The system delay is about 2.75 s when the IQL scheme is adopted, and about 2.8 s when the QMIX scheme is adopted, both higher than the proposed scheme. In the RD scheme, each user randomly selects the location of DT deployment, and the system delay generates certain jitter because of the random selection. Compared with the three benchmarks, our method can select the optimal deployment strategy based on the multi-agent network framework we trained, according to different users and network states.

\begin{figure}[htb]
	\centering
	\includegraphics[scale=0.36]{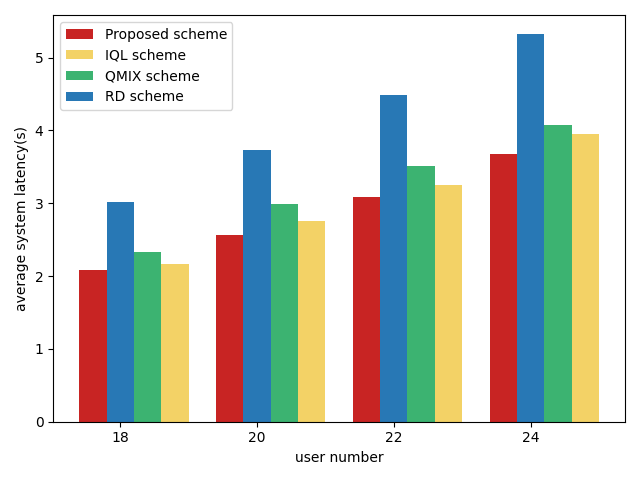}
	\caption{Numbers of users vs average system latency for different algorithms.} \label{fig:Prediction}
\end{figure}

Fig. 6 shows the comparative average system delays across four distinct schemes, each assessed under varying user counts. In this simulation experiment, we specially configure parameter of users number from 18 to 24, the step size is 2, and the other parameters are based on the default configuration mentioned above. With the increased of users number, the average system delay increases in all schemes. Because more users will make more service requests, thus occupying more resources and reducing the average quality of services for users. With different numbers of users, the proposed scheme always achieves the lowest average system delay compared with the other three benchmarks.

\begin{figure}[htb]
	\centering
	\includegraphics[scale=0.36]{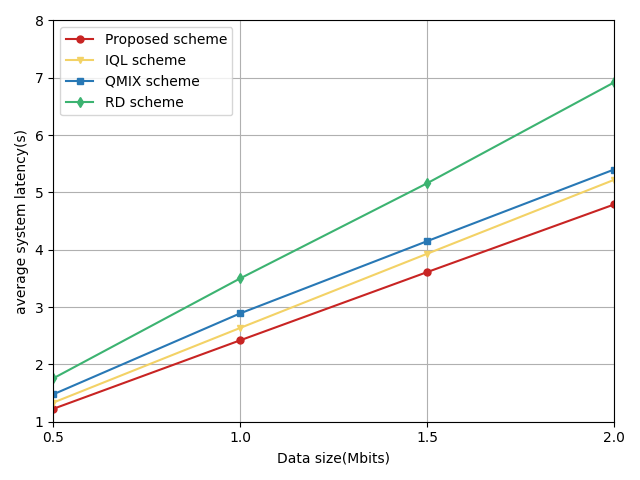}
	\caption{Data size vs average system latency for different algorithms.} \label{fig:Prediction}
\end{figure}

In Fig. 7, we compare the average system delay of different schemes with different parameters. We gradually increase the user data size from 0.5 to 2. As the amount of user data increases, the average system latency increases for all scheme. The reason is that larger user data leads to the increased time required to transmit and process data. With different user data sizes, the proposed scheme achieves the best effect. The scheme we adopted has good robustness, and can adapt to different network environments to give the optimal strategy.

\section{Conclusion}

We propose a new DT network model based on a STIN, we first propose a DT-driven STIN system model, including users, compute nodes, satellites, and cloud servers. We propose the multi-layer deployment of DTs, aiming to reduce system latency and meet user service requirements and alleviating the lack of flexibility of a traditional DITEN. In the case of dynamic network changes, we use a MARL algorithm to solve the proposed multi-layer deployment problem. Through comprehensive simulation encompassing diverse network conditions, it has been validated that the proposed approach significantly reduce system latency.

\bibliographystyle{unsrt}  
\bibliography{references}

\begin{thebibliography}{10}

\bibitem{1}
Zhen~Jiang Zhang, Wen~Yu Zhang, and Fan-Hsun Tseng.
\newblock Satellite mobile edge computing: Improving qos of high-speed satellite-terrestrial networks using edge computing techniques.
\newblock {\em IEEE Network}, 33(1):70--76, 2019.

\bibitem{2}
Nei Kato, Zubair~Md. Fadlullah, Feng~Xiao Tang, Bomin Mao, Shigenori Tani, Atsushi Okamura, and Jiajia Liu.
\newblock Optimizing space-air-ground integrated networks by artificial intelligence.
\newblock {\em IEEE Wirel Commun}, 26(4):140--147, 2019.

\bibitem{3}
Jia~Jia Liu, Yong~Peng Shi, Zubair~Md. Fadlullah, and Nei Kato.
\newblock Space-air-ground integrated network: A survey.
\newblock {\em IEEE Commun Surv Tutorials}, 20(4):2714--2741, 2018b.

\bibitem{4}
Zheng~Quan Zhang, Yue Xiao, Zheng Ma, Ming Xiao, Zhiguo Ding, Xianfu Lei, George~K. Karagiannidis, and Pingzhi Fan.
\newblock 6g wireless networks: Vision, requirements, architecture, and key technologies.
\newblock {\em IEEE Veh Technol Mag}, 14(3):28--41, 2019.

\bibitem{5}
Qing~Qing Tang, Ze~Song Fei, Bin Li, and Zhu Han.
\newblock Computation offloading in leo satellite networks with hybrid cloud and edge computing.
\newblock {\em IEEE Internet Things J}, 8(11):9164--9176, 2021.

\bibitem{38}
Guang~Chao Wang, Sheng Zhou, Shan Zhang, Zhisheng Niu, and Xuemin Shen.
\newblock Sfc-based service provisioning for reconfigurable space-air-ground integrated networks.
\newblock {\em IEEE J Sel Areas Commun}, 38(7):1478--1489, 2020.

\bibitem{39}
Shu Fu, Jie Gao, and Lian Zhao.
\newblock Integrated resource management for terrestrial-satellite systems.
\newblock {\em IEEE Trans Veh Technol}, 69(3):3256--3266, 2020.

\bibitem{40}
Xiang~Ming Zhu, Chun~Xiao Jiang, Lin~Ling Kuang, Ning Ge, Song Guo, and Jianhua Lu.
\newblock Cooperative transmission in integrated terrestrial-satellite networks.
\newblock {\em IEEE Network}, 33(3):204--210, 2019.

\bibitem{6}
Hong~Zhi Guo, Jing~Yi Li, Jia~Jia Liu, Na~Tian, and Nei Kato.
\newblock A survey on space-air-ground-sea integrated network security in 6g.
\newblock {\em IEEE Commun Surv Tutorials}, 24(1):53--87, 2022.

\bibitem{47}
Jia~Xin Zhang, Kai~Wei Wang, Rui Li, Zhaoyang Chang, Xing Zhang, and Wenbo Wang.
\newblock Macro: Mega-constellations routing systems with multi-edge cross-domain features.
\newblock {\em IEEE Wirel Commun}, 30(6):69--76, 2023.

\bibitem{48}
Peng Wang, Jia~Xin Zhang, Xing Zhang, Zhi Yan, Barry~G. Evans, and Wenbo Wang.
\newblock Convergence of satellite and terrestrial networks: A comprehensive survey.
\newblock {\em IEEE Access}, 8:5550--5588, 2020.

\bibitem{7}
Yi~Wen Wu, Ke~Zhang, and Yan Zhang.
\newblock Digital twin networks: A survey.
\newblock {\em IEEE Internet Things J}, 8(18):13789--13804, 2021.

\bibitem{8}
Paolo Bellavista, Carlo Giannelli, Marco Mamei, Matteo Mendula, and Marco Picone.
\newblock Application-driven network-aware digital twin management in industrial edge environments.
\newblock {\em IEEE Trans Ind Inf}, 17(11):7791--7801, 2021.

\bibitem{9}
Barbara~Rita Barricelli, Elena Casiraghi, and Daniela Fogli.
\newblock A survey on digital twin: Definitions, characteristics, applications, and design implications.
\newblock {\em IEEE Access}, 7:167653--167671, 2019.

\bibitem{10}
Fei Tao, He~Zhang, Ang Liu, and A.~Y.~C. Nee.
\newblock Digital twin in industry: State-of-the-art.
\newblock {\em IEEE Trans Ind Inf}, 15(4):2405--2415, 2019.

\bibitem{11}
Liang Zhao, Cheng~Cheng Wang, Kang~Lian Zhao, Daniele Tarchi, Shaohua Wan, and Neeraj Kumar.
\newblock Interlink: A digital twin-assisted storage strategy for satellite-terrestrial networks.
\newblock {\em IEEE Trans Aerosp Electron Syst}, 58(5):3746--3759, 2022.

\bibitem{12}
Yu~Ke Zhou, Ran Zhang, Jiang Liu, Tao Huang, Qinqin Tang, and F.~Richard Yu.
\newblock A hierarchical digital twin network for satellite communication networks.
\newblock {\em IEEE Commun Mag}, 61(11):104--110, 2023.

\bibitem{13}
Zhi~Sheng Yin, Nan Cheng, Tom~H. Luan, Yunchao Song, and Wei Wang.
\newblock Dt-assisted multi-point symbiotic security in space-air-ground integrated networks.
\newblock {\em IEEE Trans Inf Forensics Secur}, 18:5721--5734, 2023.

\bibitem{14}
Hui~Long Fan, Jun Long, Li~Min Liu, and Zhan Yang.
\newblock Dynamic digital twin and online scheduling for contact window resources in satellite network.
\newblock {\em IEEE Trans Ind Inf}, 19(5):7217--7227, 2023.

\bibitem{15}
Xin~Yuan Jiang, Tao Zhang, and Liang Liu.
\newblock Research on satellite qos routing algorithm based on digital twin.
\newblock In {\em Proc $11^{th}$ Int Conf on Intelligent Computing and Wireless Optical Communications}, pages 118--122, 2023.

\bibitem{16}
Zhe Ji, Sheng Wu, and Chun~Xiao Jiang.
\newblock Cooperative multi-agent deep reinforcement learning for computation offloading in digital twin satellite edge networks.
\newblock {\em IEEE J Sel Areas Commun}, 41(11):3414--3429, 2023.

\bibitem{17}
Feng~Xiao Tang, Xue~Han Chen, Tiago~Koketsu Rodrigues, Ming Zhao, and Nei Kato.
\newblock Survey on digital twin edge networks (diten) toward 6g.
\newblock {\em IEEE Open J Commun Soc}, 3:1360--1381, 2022.

\bibitem{18}
Jia~Jia Liu, Yong~Peng Shi, Lei Zhao, Yurui Cao, Wen Sun, and Nei Kato.
\newblock Joint placement of controllers and gateways in sdn-enabled 5g-satellite integrated network.
\newblock {\em IEEE J Sel Areas Commun}, 36(2):221--232, 2018a.

\bibitem{19}
Yu~Rui Cao, Hong~Zhi Guo, Jia~Jia Liu, and Nei Kato.
\newblock Optimal satellite gateway placement in space-ground integrated networks.
\newblock {\em IEEE Network}, 32(5):32--37, 2018.

\bibitem{20}
Yun~Long Lu, Xiao~Hong Huang, Ke~Zhang, Sabita Maharjan, and Yan Zhang.
\newblock Low-latency federated learning and blockchain for edge association in digital twin empowered 6g networks.
\newblock {\em IEEE Trans Ind Inf}, 17(7):5098--5107, 2021b.

\bibitem{21}
Wen Sun, Hai~Bin Zhang, Rong Wang, and Yan Zhang.
\newblock Reducing offloading latency for digital twin edge networks in 6g.
\newblock {\em IEEE Trans Veh Technol}, 69(10):12240--12251, 2020.

\bibitem{22}
Rui Dong, Changyang She, Wibowo Hardjawana, Yonghui Li, and Branka Vucetic.
\newblock Deep learning for hybrid 5g services in mobile edge computing systems: Learn from a digital twin.
\newblock {\em IEEE Trans Wirel Commun}, 18(10):4692--4707, 2019.

\bibitem{23}
Ke~Zhang, Jia~Yu Cao, and Yan Zhang.
\newblock Adaptive digital twin and multiagent deep reinforcement learning for vehicular edge computing and networks.
\newblock {\em IEEE Trans Ind Inf}, 18(2):1405--1413, 2022.

\bibitem{24}
Tong Liu, Lun Tang, Wei~Li Wang, Qianbin Chen, and Xiaoping Zeng.
\newblock Digital-twin-assisted task offloading based on edge collaboration in the digital twin edge network.
\newblock {\em IEEE Internet Things J}, 9(2):1427--1444, 2022.

\bibitem{25}
Qi~Guo, Feng~Xiao Tang, and Nei Kato.
\newblock Resource allocation for aerial assisted digital twin edge mobile network.
\newblock {\em IEEE J Sel Areas Commun}, 41(10):3070--3079, 2023.

\bibitem{42}
Yi~Long Hui, Yi~Qiu, Zhou Su, Zhisheng Yin, Tom~H. Luan, and Khalid Aldubaikhy.
\newblock Digital twins for intelligent space-air-ground integrated vehicular network: Challenges and solutions.
\newblock {\em IEEE Internet Things Mag}, 6(3):70--76, 2023.

\bibitem{43}
Bo~Min Mao, Xue~Ming Zhou, Jia~Jia Liu, and Nei Kato.
\newblock Digital twin satellite networks toward 6g: Motivations, challenges, and future perspectives.
\newblock {\em IEEE Network}, 38(1):54--60, 2024.

\bibitem{44}
Trung~Q. Duong, Long~D. Nguyen, Tinh~T. Bui, Khanh~D. Pham, and George~K. Karagiannidis.
\newblock Machine learning-aided real-time optimized multibeam for 6g integrated satellite-terrestrial networks: Global coverage for mobile services.
\newblock {\em IEEE Network}, 37(2):86--93, 2023.

\bibitem{45}
Qi~Guo, Feng~Xiao Tang, Tiago~Koketsu Rodrigues, and Nei Kato.
\newblock Five disruptive technologies in 6g to support digital twin networks.
\newblock {\em IEEE Wirel Commun}, 31(1):149--155, 2024.

\bibitem{26}
Yun~Long Lu, Sabita Maharjan, and Yan Zhang.
\newblock Adaptive edge association for wireless digital twin networks in 6g.
\newblock {\em IEEE Internet Things J}, 8(22):16219--16230, 2021.

\bibitem{27}
Hui Zhang, Tian~Xiang Luo, and Qian~Qian Wang.
\newblock Adaptive digital twin server deployment for dynamic edge networks in iot system.
\newblock In {\em IEEE/CIC Int Conf on Communications in China}, pages 1--6, 2023.

\bibitem{28}
Olga Chukhno, Nadezhda Chukhno, Giuseppe Araniti, Claudia Campolo, Antonio Iera, and Antonella Molinaro.
\newblock Placement of social digital twins at the edge for beyond 5g iot networks.
\newblock {\em IEEE Internet Things J}, 9(23):23927--23940, 2022.

\bibitem{29}
Ya~Dong Zhang, Hai~Bin Zhang, Yun~Long Lu, Wen Sun, Lan Wei, Yan Zhang, and Bin Wang.
\newblock Adaptive digital twin placement and transfer in wireless computing power network.
\newblock {\em IEEE Internet Things J}, 11(6):10924--10936, 2024.

\bibitem{30}
Mehrad Vaezi, Kiana Noroozi, Terence~D. Todd, Dongmei Zhao, and George Karakostas.
\newblock Digital twin placement for minimum application request delay with data age targets.
\newblock {\em IEEE Internet Things J}, 10(13):11547--11557, 2023.

\bibitem{31}
Zhi~Xiu Yao, Shi~Chao Xia, Yun Li, and Guangfu Wu.
\newblock Cooperative task offloading and service caching for digital twin edge networks: A graph attention multi-agent reinforcement learning approach.
\newblock {\em IEEE J Sel Areas Commun}, 41(11):3401--3413, 2023.

\bibitem{41}
Xiao~Nan Li, Hai~Jun Zhang, Huan Zhou, Ning Wang, Keping Long, Saba Al-Rubaye, and George~K. Karagiannidis.
\newblock Multi-agent drl for resource allocation and cache design in terrestrial-satellite networks.
\newblock {\em IEEE Trans Wirel Commun}, 22(8):5031--5042, 2023.

\bibitem{33}
Zi~Yang Guo, Zhen~Yu Chen, Peng Liu, Jianjun Luo, Xun Yang, and Xinghua Sun.
\newblock Multi-agent reinforcement learning-based distributed channel access for next generation wireless networks.
\newblock {\em IEEE J Sel Areas Commun}, 40(5):1587--1599, 2022.

\bibitem{35}
Xiang Tan, Li~Zhou, Hai~Jun Wang, Yuli Sun, Haitao Zhao, Boon-Chong Seet, Jibo Wei, and Victor C.~M. Leung.
\newblock Cooperative multi-agent reinforcement-learning-based distributed dynamic spectrum access in cognitive radio networks.
\newblock {\em IEEE Internet Things J}, 9(19):19477--19488, 2022.

\bibitem{34}
Klaus Greff, Rupesh~K. Srivastava, Jan Koutník, Bas~R. Steunebrink, and Jürgen Schmidhuber.
\newblock Lstm: A search space odyssey.
\newblock {\em IEEE Trans Neural Networks Learn Syst}, 28(10):2222--2232, 2017.

\bibitem{32}
Peter Sunehag, Guy Lever, Audrunas Gruslys, Wojciech~Marian Czarnecki, Vinicius Zambaldi, Max Jaderberg, Marc Lanctot, Nicolas Sonnerat, Joel~Z. Leibo, Karl Tuyls, and Thore Graepel.
\newblock Value-decomposition networks for cooperative multi-agent learning based on team reward.
\newblock In {\em Proc $17^{th}$ Int Conf on Autonomous Agents and MultiAgent Systems}, AAMAS '18, page 2085–2087, Richland, SC, 2018. International Foundation for Autonomous Agents and Multiagent Systems.

\bibitem{37}
Ming Tang and Vincent W~S Wong.
\newblock Deep reinforcement learning for task offloading in mobile edge computing systems.
\newblock {\em IEEE Trans Mobile Comput}, 21(6):1985--1997, 2022.

\bibitem{36}
Xiang Tan, Li~Zhou, Haijun Wang, Yuli Sun, Haitao Zhao, Boon-Chong Seet, Jibo Wei, and Victor C.~M. Leung.
\newblock Cooperative multi-agent reinforcement-learning-based distributed dynamic spectrum access in cognitive radio networks.
\newblock {\em IEEE Internet of Things Journal}, 9(19):19477--19488, 2022.

\end{thebibliography}

\end{document}